\documentclass[aps,twocolumn,showpacs,pra]{revtex4-1}
\usepackage{graphicx,color}
\usepackage{hyperref}
\begin{document}
\title{Majorana representation, qutrit Hilbert 
space and NMR implementation of qutrit gates}
\author{Shruti Dogra}
\email{shrutidogra@iisermohali.ac.in}
\address{Department of Physical Sciences,
Indian Institute of Science Education \&
Research Mohali, Sector 81 SAS Nagar,
Manauli PO 140306 Punjab India.}
\author{Kavita Dorai}
\email{kavita@iisermohali.ac.in}
\address{Department of Physical Sciences,
Indian Institute of Science Education \&
Research Mohali, Sector 81 SAS Nagar,
Manauli PO 140306 Punjab India.}
\author{Arvind}
\email{arvind@iisermohali.ac.in}
\address{Department of Physical Sciences,
Indian Institute of Science Education \&
Research Mohali, Sector 81 SAS Nagar,
Manauli PO 140306 Punjab India.}
\begin{abstract}
We report a study of the Majorana geometrical representation
of a qutrit, where a pair of points on a unit sphere
represents its quantum states.  A canonical form for qutrit
states is presented, where every state can be obtained from
a one-parameter family of states via $SO(3)$ action. The
notion of spin-1 magnetization which is invariant under
$SO(3)$ is geometrically interpreted on the Majorana sphere.
Furthermore, we describe the action of several quantum gates
in the Majorana picture and experimentally implement these
gates on a spin-1 system (an NMR qutrit) oriented in a
liquid crystalline environment.  We study the dynamics of
the pair of points representing a  qutrit state under
various useful quantum operations and connect them to
different NMR operations.  Finally, using the Gell Mann
matrix picture we experimentally implement a scheme for
complete qutrit state tomography.
\end{abstract}
\maketitle
\section{Introduction}
\label{introduction}
The Bloch sphere provides a representation of the
quantum states of a single qubit onto
$\mathcal{S}^2$(a unit sphere  in three real
dimensions), with pure states mapped onto the
surface and the mixed states lying in the
interior~\cite{nielsen-book-02,bengtsson-book-06}.
This geometrical representation is  useful in
providing a visualization of  quantum states and
their transformations, particularly in the case of
NMR-based quantum computation, where the
spin-$\frac{1}{2}$ magnetization and its
transformation through NMR rf pulses is visualized
on the Bloch sphere~\cite{levitt-book-2008}.
There have been  several proposals for the
geometrical representation for higher-level
quantum
systems~\cite{arvind-jpa-1997,goyal-qph-2011,ashourisheikhi-ijqi-2013,
messina-pra-2010,makela-ps-2010,sen-ap-2012,planat-geom-2011,yan-pra-2010}
however,  extensions of a Bloch sphere-like
picture to higher spins is not straightforward.  A
geometrical representation was proposed by
Majorana in which, a pure state of a spin `$s$' is
represented by `$2s$' points on the surface of a
unit sphere, called the Majorana
sphere~\cite{majorana-nc-1932}.  The Majorana
representation for spin$-s$ systems has found
widespread applications such as determining
geometric phase of
spins~\cite{hannay-jpa-1998,liu-qph-2014},
representing $N$ spinors by $N$
points~\cite{devi-qip-2012}, geometrical
representation of multi-qubit entangled
states~\cite{zyczkowski-pra-2012,devi-qph-2010,devi-qph-2010-2},
statistics of chaotic quantum dynamical
systems~\cite{hannay-jpa-1996} and characterizing
polarized light~\cite{hannay-jmo-1998}.

A single qutrit (three-level quantum system) 
is of particular importance 
in qudit-based ($d$-level
quantum system)  
quantum computing
schemes~\cite{li-sr-2013,zobov-jetp-2011,akyuz-oc-2008,terriza-prl-2004,bruss-prl-2002,langford-prl-2004,lanyon-prl-2008}.
A qutrit
is the smallest system that exhibits inherent quantum
features such as
contextuality~\cite{peres-jpa-1991,yu-prl-2012,
kurzynski-pra-2012}, which has been conjectured
to be a resource for quantum
computing~\cite{ladd-nature-2010,dogra-pla-2014,nature-2014,gedik-scirep-2015,prl-2017}.
NMR qudit quantum computing can be performed by using nuclei
with spin $s > \frac{1}{2}$ or can be modeled by two or more coupled
spin-$\frac{1}{2}$ nuclei~\cite{gopinath-pra-2006,das-ijqi-2003}.

In this work we use the Majorana sphere
description of a single qutrit, where states of a
qutrit are represented by a pair of points on a
unit sphere, to provide insights into the qutrit
state space. The group of unitary transformations
$SU(3)$ and the group of rotations in three
dimensions $SO(3)$ play an important role in the
study of a qutrit.  We propose a way to
parameterize single-qutrit states, where arbitrary
qutrit states are generated from a one-parameter
family of a canonical subset of states via the
action of $SO(3)$ transformations.  The unitary
transformation corresponding to the physical
rotations represented by $SO(3)$  act rigidly on
the Majorana sphere.  The $SO(3)$ invariant notion
of spin magnetization finds a natural
representation on the Majorana sphere.  States are
identified as ``pointing'' or ``non-pointing'',
depending upon the zero or non-zero value of the
spin magnetization.  As the state of a qutrit
evolves, the pair of points representing the state
moves on the surface of the Majorana sphere.  The
physical rotations represented by $SO(3)$ when
implemented unitarily, and which correspond to
rigid rotations of the pair of points on the
Majorana sphere are experimentally realized in NMR
via spin-selective pulses.  The transformation
generated by the standard $SU(3)$ generators given
by $\Lambda$-matrices~\cite{arvind-jpa-1997}
correspond in most cases 
to the movement of only one point on
the Majorana sphere, and are experimentally
realized in NMR via transition-selective pulses.

The above correspondences are used to visualize
and experimentally implement different gates on a
qutrit that are used in NMR quantum computing,
namely SWAP gates, controlled-phase shift gates
and a three-dimensional analogue of the Hadamard
gate.  We used a spin-1 deuterium nucleus of a
chloroform-D molecule to realize an NMR qutrit.
In order to obtain a well resolved pair of
single-quantum transitions in the deuterium NMR
spectrum, the chloroform-D molecule is embedded in
an anisotropic liquid crystalline environment.
The experimental implementation of these ternary
quantum gates is validated by complete quantum
state tomography which we carry out using
the Gell Mann matrices.

The material in this paper is arranged as follows:
Section~\ref{scheme} contains a discussion of the
Majorana representation of a spin-1 particle, while
section~\ref{so3rotations} focuses on the dynamics of a
spin-1 particle under $SU(3)$ and $SO(3)$ operations.
The canonical state transformations
and the Majorana
representation of the spin-1 magnetization vector 
are described in
sections~\ref{canonical-state} and
\ref{magnetization-vector}, respectively.
Section~\ref{computing}
describes the basic requirements to perform NMR quantum
computing using a single qutrit.
The qutrit system, initialization scheme, 
quantum gate realizations and
experimental NMR
implementations of qutrit gates
are described in sections~\ref{qutrit}~-~\ref{qgates},
respectively.  Section~\ref{conclusions}
contains a few concluding remarks.
\section{Majorana representation} 
\label{scheme}
In the Majorana representation each state of a
spin-$s$ quantum system is represented by $2s$
points on the surface of a unit
sphere~\cite{majorana-nc-1932}.  Consider the
$2s+1$ orthonormal eigen vectors of the angular
momentum operator $L_z$, $\{\vert j \rangle\}$
with $j$ ranging from $-s$ to $+s$, providing a
basis for the Hilbert space of spin $s$.  The most
general quantum pure state for the system is
given in terms of $2s+1$ complex coefficients
$C_j$ as
\begin{equation}
\vert {\Psi} \rangle = \sum_{j=-s}^{+s} C_{j}\vert j \rangle 
= C_{-s} \vert{-s} \rangle \cdots + C_{0} \vert{0} \rangle  
\cdots + C_{+s} \vert{+s} \rangle 
\label{expansion}
\end{equation}
with $\sum_{j=-s}^s \vert C_j \vert^2=1$.
\begin{figure}[h!]
\centering
\includegraphics[scale=1]{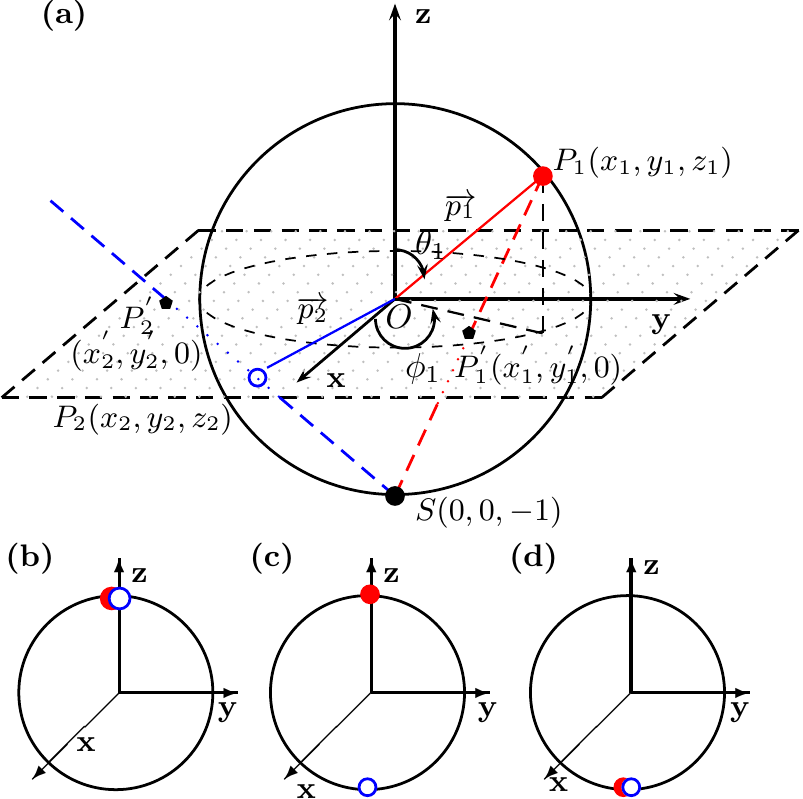}
\caption{A qutrit on the Majorana sphere is represented by two 
points $P_1$ and $P_2$, connected with the center of the 
sphere by lines shown in red and blue respectively.
$\theta_1$, $\phi_1$ are the polar and azimuthal angles 
corresponding to point $P_{1}$ 
($\theta_2,
\phi_2$ are the angles for point $P_2$).
(a) Roots of the Majorana polynomial are shown in the plane $z=0$ 
by points $P_1^{\prime}$ and $P_2^{\prime}$, whose stereographic projection
give rise to the Majorana representation.
Three examples are shown corresponding to the Majorana 
representation of single-qutrit basis vectors
(b) $\vert +1 \rangle$, (c) $\vert 0 \rangle$ and 
(d) $\vert -1 \rangle$.  One of
the points is shown as a solid (red) circle, while the other 
point is represented by an
empty (blue) circle.}
\label{example}
 \end{figure}
Majorana introduced a polynomial of degree $2s$
whose coefficients are derived from the expansion
coefficients given in Eq.~(\ref{expansion}),
\begin{eqnarray} 
&&a_{0}\zeta^{2s}+a_{1}\zeta^{2s-1}+ \dots
\dots + a_{2s}=0, \nonumber \\ && a_{r}=(-1)^{r}
\frac{C_{s-r}}{\sqrt{(2s-r)!r!}}.  
\end{eqnarray} 
This polynomial has $2s$ complex roots which are
determined by the coefficients $C_j$ and
completely determine the quantum state
of the spin $s$ system,  upto an overall phase.
These points are plotted in  the complex
$xy$-plane and the inverse stereographic projection of
these roots with respect to the south pole gives
us $2s$ points on the surface of a unit sphere.
These $2s$ points provide the Majorana
representation of the state $\vert \psi \rangle$.

For the specific case of a qutrit that interests
us, the three-dimensional Hilbert space is
spanned by the three eigen vectors $\{ \vert -1
\rangle, \vert 0\rangle, \vert +1 \rangle \}$ of
the angular momentum operator $\Sigma_3$ ($L_z$).
The general state is given by
\begin{equation}
\vert{\Psi} \rangle = C_{-1} \vert{-1} \rangle + C_{0}
\vert{0} \rangle + C_{+1} \vert{+1} \rangle
\end{equation}
and the corresponding Majorana polynomial is a
quadratic equation. Fig.~\ref{example}(a) shows
the Majorana sphere representation of a qutrit,
with the roots of the Majorana polynomial shown as
points $P_1^{\prime}(x_1^{\prime},y_1^{\prime},0)$
and $P_2^{\prime}(x_2^{\prime},y_2^{\prime},0)$.
The corresponding points obtained via 
inverse stereographic projections with
respect to the south pole are shown as  points
$P_1$ and $P_2$, and they determine the quantum state
of a qutrit system in the Majorana representation.
The three basis states are shown on
the Majorana sphere in Figs.~\ref{example}(b)-(d)
respectively. The Majorana representation of
$\vert +1 \rangle$ consists of the points
$(0,0,1)$ and $(0,0,1)$ which are overlapping
points on the north pole. Similarly, for the state
$\vert -1 \rangle$, both the points lie on the
south pole. For the state $\vert 0 \rangle$, one
point lies on the north pole while the other point
lies on the south pole.  In general, the points
$P1$ and $P_2$ can lie anywhere on the sphere and
we can find such a representation for every pure
quantum state of a qutrit.

Let us consider two arbitrary points
$P_1(\theta_1, \phi_1)$ and $P_2(\theta_2,
\phi_2)$ on the Majorana sphere with $\theta_i \in
[0,\pi]$ and $\phi_i \in [0,2\pi]$.
Fig.~\ref{example}(a) shows these points as well
as their stereographic projections from the south
pole.  The corresponding  points $P_1^{\prime}$
and $P_2^{\prime}$ in the complex plane are also
shown.  The coordinates of these points $P_i^{'}$,
$i = 1,2$ in the complex plane in terms of the
polar coordinates of $P_1$ and $P_2$ are given by
\begin{eqnarray}
x_i^{'}&=&(\sin \theta_i \cos \phi_i)/(1+\cos
\theta_i) \nonumber \\
y_i^{'}&=&(\sin \theta_i \sin \phi_i)/(1+\cos \theta_i)
\label{general_majorana}
\end{eqnarray}
Thus the two roots of the second degree Majorana
polynomial for a qutrit are $e^{\iota
\phi_i}\tan\frac{\theta_i}{2}$, $i=1,2$. From the
roots we find the  coefficients of the quadratic
Majorana polynomial and the most general qutrit
state:
\begin{eqnarray}
\vert \psi_g \rangle&=& \Gamma \left (
\begin{array}{c} \sqrt{2} \cos \frac{\theta_1}{2}
\cos \frac{\theta_2}{2} \\ e^{\iota \phi_1} \sin
\frac{\theta_1}{2} \cos \frac{\theta_2}{2}
+e^{\iota \phi_2} \cos \frac{\theta_1}{2} \sin
\frac{\theta_2}{2} \\ \sqrt{2} e^{\iota
(\phi_1+\phi_2)} \sin \frac{\theta_1}{2} \sin
\frac{\theta_2}{2}
\end{array}\right) 
\nonumber \\ 
\Gamma&=&\sqrt{2}\left[3+\cos \theta_1 \cos \theta_2 + 
\sin \theta_1 \sin \theta_2
\cos(\phi_1-\phi_2)\right]^{-\frac{1}{2}} 
\nonumber \\
&&\theta_i \in [0,\pi], \quad \quad
\phi_i \in [0,2\pi], \quad i=1,2.  \nonumber \\
\label{genstate}
\end{eqnarray}
As evident from Eq.~(\ref{genstate}), a pure
state of a single qutrit is represented by six
real (three complex) parameters; using one
normalization condition reduces this number to
five and with the states being defined upto an
overall phase, the number of parameters determining
a general pure qutrit state reduces to four.
\section{$SU(3)$ and $SO(3)$ canonical states and
invariant magnetization}
\label{so3rotations}
The group of unitary transformation in three
dimensions $SU(3)$ and the group of  rotations in
three dimensions $SO(3)$ (embedded in $SU(3)$ as a
subgroup) play an important role in the study of a
qutrit.  A convenient set of generators for the
group  $SU(3)$ is provided by the Lambda matrices
due to Gell Mann.
\begin{eqnarray} 
&&
\Lambda_{1} =\left(
\begin{array}{ccc}0 & 1 & 0 \\ 1 & 0 & 0  \\ 0 & 0
& 0 \end{array}  \right),
\Lambda_{2}
=\left( \begin{array}{ccc}0 & -\iota & 0 \\ \iota
& 0 & 0  \\ 0 & 0 & 0 \end{array}  \right),
\Lambda_{3} =\left(
\begin{array}{ccc}1 & 0 & 0 \\ 0 & -1 & 0  \\ 0 &
0 & 0 \end{array}  \right), 
\nonumber \\
&&
\Lambda_{4} =\left( \begin{array}{ccc}0 & 0 & 1 \\
0 & 0 & 0  \\ 1 & 0 & 0 \end{array}  \right),
\Lambda_{5} =\left( \begin{array}{ccc}0 & 0 &
-\iota \\ 0 & 0 & 0  \\ \iota & 0 & 0 \end{array}
\right),
\Lambda_{6} =\left(
\begin{array}{ccc}0 & 0 & 0 \\ 0 & 0 & 1  \\ 0 & 1
& 0 \end{array}  \right),
\nonumber \\
&&
\Lambda_{7}
=\left( \begin{array}{ccc}0 & 0 & 0 \\ 0 & 0 &
-\iota  \\ 0 & \iota & 0 \end{array}  \right),
\quad
\Lambda_{8} =
\frac{1}{\sqrt{3}}\left( \begin{array}{ccc}1 & 0 &
0 \\ 0 & 1 & 0  \\ 0 & 0 & -2 \end{array}
\right).
\end{eqnarray}
The transformations that these
matrices generate upon exponentiation are given by
\begin{equation}
U_{\Lambda_i}=e^{\iota \theta
\Lambda i}
\end{equation}
and they target a two-level subspace of the
three-level quantum system. 
These operations can be experimentally implemented 
via transition-selective pulses in NMR,
which are governed 
by the single transition operators~\cite{wokaun-jcp-1977},
generating transformations:
\begin{equation}
e^{\iota \xi I_{k}^{rs}} = I - I^{rs} + 
\cos\frac{\xi}{2} \, I_{k}^{rs} + 
2 \iota \sin\frac{\xi}{2} \, I_{k}^{rs}
\end{equation}
which are unitary operators exciting transitions between levels
$r,s$ about the $k^{th}$ axis by an angle $\xi$.
$I_{k}^{rs}$ is the product operator element 
corresponding to the $(r,s)$ subspace 
and $k \in \{ x,y,z \}$. 
A direct connection between single-transition operators in
NMR and the Gell Mann matrices can be established:
\begin{eqnarray}
I_x^{12} &=& \frac{1}{2} \Lambda_1, \quad I_y^{12} = 
\frac{1}{2} \Lambda_2, 
\quad I_z^{12} = \frac{1}{2} \Lambda_3 
\nonumber \\
I_x^{23} &=& \frac{1}{2} \Lambda_6, \quad I_y^{23} = 
\frac{1}{2} \Lambda_7, \quad 
I_z^{23} = \frac{1}{2} 
\left(\sqrt{3}\Lambda_8-\Lambda_3 \right)
\nonumber \\
I_x^{13} &=& \frac{1}{2} 
\Lambda_4, \quad I_y^{13} = \frac{1}{2} \Lambda_5, \quad
I_z^{13} = \frac{1}{2} 
\left(\sqrt{3}\Lambda_8+\Lambda_3 \right)
\nonumber \\
\label{n1-n3} 
\end{eqnarray}
The above connection leads to the possibility of
implementing transformations generated by Gell
Mann matrices via rf pulses on single transitions.
The transformations corresponding to the
indices ($13$) in Equation~(\ref{n1-n3}) are associated
with a double-quantum transition and cannot be
implemented by one rf pulse on a single
transition and instead involve cascades of
pulses~\cite{dorai-jmr-1995}.

The group $SO(3)$ plays a very important role in
qutrit physics: on the one hand it acts as
the group of rotations in three dimensions and on
the other it is a subgroup of $SU(3)$ through its
spin 1 unitary representation.
The generators of this group in its defining
representation  denoted by $J_j$ and in its 
three dimensional unitary representation denoted by
$\Sigma_j$ are given by
\begin{eqnarray}
J_1&=i\left(\begin{array}{ccc} 
0&0&0\\0&0&1\\0&-1&0
\end{array}\right), \quad
\Sigma_1 &= 
\frac{1}{\sqrt{2}} 
\left( \begin{array}{ccc}
0 & 1 & 0 \\
1 & 0 & 1 \\
0 & 1 & 0 \\
\end{array}  \right),\nonumber \\
J_2&=i \left(\begin{array}{ccc} 
0&0&1\\0&0&0\\-1&0&0
\end{array}\right),  \quad
\Sigma_2 &= 
\frac{1}{\sqrt{2}} 
\left( \begin{array}{ccc}
0 & -\iota & 0 \\
\iota & 0 & -\iota \\
0 & \iota & 0 \\
\end{array}  \right), \nonumber \\
J_3&=i\left(\begin{array}{ccc}
0&-1&0\\1&0&0\\0&0&0\end{array}
\right),\quad
\Sigma_3 &=
\left( \begin{array}{ccc}
1 & 0 & 0 \\
0 & 0 & 0 \\
0 & 0 & -1 \\
\end{array}  \right)
\end{eqnarray}
The generators $J_j$ generate finite rotations about
the $x,y$ and $z$ axes respectively while the generators
$\Sigma_j$ generate the corresponding three
dimensional unitary transformations:
\begin{equation}
R_j(\xi)=e^{i\xi J_i}\in SO(3);\,\,
U_j(\xi)=e^{i\xi
\Sigma_i}\in SO(3)\subset SU(3)
\end{equation}
When the group $SO(3)$ acts on the real physical
space via its defining representation and the
corresponding action on quantum states is via its
unitary representation, the points on the Majorana
sphere transform rigidly and as per the defining
representation of $SO(3)$. This is a very useful
feature and can be exploited in describing and
transforming quantum states of the qutrit.

Consider a single-qutrit state, 
$\vert \psi_{initial} \rangle$ 
whose Majorana representation is given as points 
$P_{1}(x_1,y_1,z_1)$
and $P_{2}(x_2,y_2,z_2)$ (Fig.~\ref{example}(a)).
Joining each of these points with the center of 
the sphere ($O(0,0,0)$), one obtains unit vectors
$\overrightarrow{OP_1}$ and $\overrightarrow{OP_2}$,
which are represented as
$\overrightarrow{p_1}$ and $\overrightarrow{p_2}$
respectively in Fig.~\ref{example}(a). Under the
$SO(3)$ action these vectors rotate together and
rigidly while the
quantum state in the Hilbert space undergoes the
corresponding unitary transformation 
\begin{equation}
U_{i}(\xi)=e^{\iota \xi \Sigma_{i}}
=I + (\cos\xi - 1) \;
\Sigma_{i}^{2} + \iota \sin \xi \; \Sigma_{i} 
\end{equation}
where $I$ is the $3 \times 3$ identity matrix.
When $U_i(\xi)$ acts on a single-qutrit state 
$\vert \psi \rangle$
in the $3-$dimensional Hilbert space, it leads to
the rotation of the two points
representing a qutrit on the Majorana
sphere by angle $\xi$ about $i$-axis.

To illustrate this  
let us consider a general single-qutrit state
($\vert \psi_g \rangle$) given in
Eq.~(\ref{genstate}) as the initial state ($\vert
\psi_{initial} \rangle$).  The Majorana
representation of this state is given by two
points: $P_1(\sin \theta_1 \cos \phi_1, \sin
\theta_1 \sin \phi_1,\cos \theta_1)$ and $P_2(\sin
\theta_2 \cos \phi_2, \sin \theta_2 \sin
\phi_2,\cos \theta_2)$.
Considering an operator $U_z(\xi_z=\phi_1)$, under
whose action the general single-qutrit state
gives rise to the final state: 
\begin{eqnarray}
\vert \psi_{final} \rangle= \Gamma \left (
\begin{array}{c} \sqrt{2} \cos \frac{\theta_1}{2}
\cos \frac{\theta_2}{2} \\ \sin \frac{\theta_1}{2}
\cos \frac{\theta_2}{2} +e^{\iota (\phi_2-\phi_1)}
\cos \frac{\theta_1}{2} \sin \frac{\theta_2}{2} \\
\sqrt{2} e^{\iota (\phi_2-\phi_1)} \sin
\frac{\theta_1}{2} \sin \frac{\theta_2}{2}
\end{array}\right) \nonumber \\
\Gamma=\frac{\sqrt{2}}{\sqrt{3+\cos \theta_1 \cos
\theta_2 + \sin \theta_1 \sin \theta_2
\cos(\phi_1-\phi_2)}}. \nonumber \\ \end{eqnarray}
Majorana representation of state $\vert
\psi_{final} \rangle$ is given by the  points:
$Q_1\left( \sin \theta_1,0,\cos \theta_1\right)$
and $Q_2\left( \sin \theta_2 \cos (\phi_2-\phi_1),
\sin \theta_2 \sin (\phi_2-\phi_1),\cos \theta_2
\right)$. These points are the same as those obtained
from ($P_1, P_2$) by the direct action of the
group $SO(3)$ via its defining representation.
Interestingly, the $SO(3)$ transformations
on a single qutrit are
experimentally implemented by a non-selective rf
pulse in NMR which  is in contrast to the
transformations generated by the Gell Mann
matrices where single transition pulses are
involved. 
\subsection{Canonical states} 
\label{canonical-state} 
\begin{figure}[h!]
\includegraphics{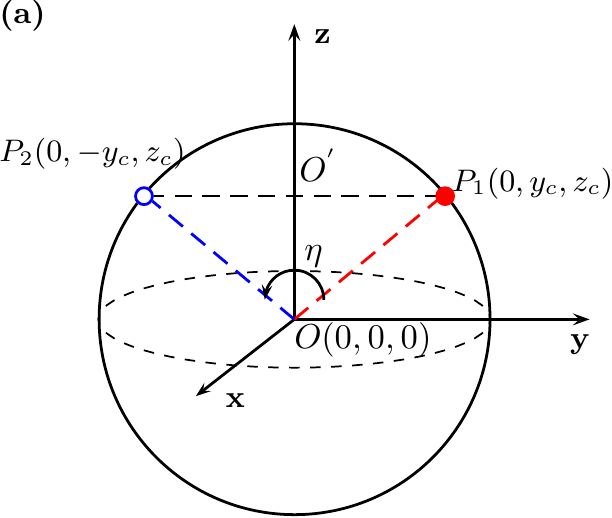}
\caption{Majorana representation of canonical state,
 $\vert \psi_c \rangle$ shown as two points $P_1$ and 
$P_2$ in red and blue colors respectively. Dotted lines 
joining $P_1$ and $P_2$ with center $O(0,0,0)$, enclose 
angle $\eta$ at the center of the sphere.
}
\label{canonical-rep}
\end{figure}

Can the $SO(3)$ transformations described above be
used to develop a convenient parameterization of
the qutrit states?  It turns out that we can
obtain all the states of a qutrit from a one-parameter 
family of states via the $SO(3)$ action.
Consider a one-parameter family  of states
described by a real parameter $\alpha$:
\begin{equation}
 \left\vert{\psi_c}\right\rangle=
\left( \begin{array}{c}
        \sin \alpha \\ 0 \\ \cos \alpha
     \end{array}
\right)
\qquad {\rm where} \; \alpha \in [0,\pi/2]
\label{canonical_state_eq}
\end{equation}
Majorana representation of 
$\left\vert{\psi_c}\right\rangle$ consists  of two points,
$P_1(0,y_c,z_c)$ and $P_2(0,-y_c,z_c)$ where
$y_c=\frac{\displaystyle \sqrt{2\sin
2\alpha}}{\displaystyle \sin \alpha + \cos \alpha}$
and $z_c=\frac{\displaystyle \sin \alpha - \cos
\alpha}{\displaystyle \sin \alpha + 
\cos \alpha}$.
These points lie on the great circle in the plane $x=0$, such
that they have same value of $z-$ coordinates and equal and
opposite values of the projection on $y-$axis as shown in 
Fig.~\ref{canonical-rep} and the
angle enclosed between the points $P_1$ and $P_2$ at the 
center of sphere is $\eta = 2\sin^{-1}y_c$.
As $\alpha$ varies from $0 \leq \alpha \leq
\frac{\pi}{2}$, $\eta$ assumes values $0 \leq \eta
\leq 2\pi$, thus the two points $P_1$ and $P_{2}$
can lie anywhere on the great circle in the plane $x=0$.
$OO^{'}$ is the perpendicular bisector of line
joining the points $P_1$, $P_2$ and divides the
angle
$\eta$ into half. Under the action of $U_i(\xi_i)$
transformation on spin-1 quantum state,
corresponding $OO^{'}$ in the Majorana
representation rotates like an ordinary vector in
coordinate space.  Depending upon the value of
$\alpha$ the magnitude of bisector
$OO^{'}$ attains values ranging from $+1$ to $-1$
while this magnitude remains invariant under
rotations. This opens up an interesting
possibility of obtaining a family of states
related to each other by $SO(3)$ transformations
for each value of the magnitude of $OO^{'}$.  It
turns out that  each and every state of a single
qutrit can be obtained from this canonical state
using $SO(3)$ transformations.

To see this explicitly, we give 
a set of three $SO(3)$ operators with specific
angles and rotation axes, under whose action
the spin-1 general state can be taken to the spin-1
canonical state and vice versa.
Since we know that there exists a
one-to-one correspondence between the operators
$U_i(\xi_i)$ in Hilbert space and $R_{i}(\xi_i)$
in the coordinate space, we make use of the
Majorana representation of the states $\vert
\psi_g \rangle$, $\vert \psi_c \rangle$ (defined in
Eqs.~(\ref{genstate}) and~(\ref{canonical_state_eq}) to find
the set of rotation matrices $R_{i}(\xi_i)s$ and
then the corresponding unitary matrices.

Starting with the Majorana representation of a
general state $\vert \psi_g \rangle$ consisting of
points $P_1$ and $P_2$ as depicted in
Fig.~\ref{example}(a) with their polar coordinate
on the sphere given via
Eq.~(\ref{general_majorana}), we aim to rotate and
go to points representing the canonical state $\vert
\psi_c \rangle$ which consists of a pair of points
on the great circle in the plane $x=0$. 
This is achieved by rotating the points  
$P_1$and $P_2$ 
about $y-$axis by angle $\gamma$ followed by a
rotation 
$z-$axis by angle $\beta$ such 
that both the points can be brought to the plane 
$x=0$. Further a rotation about x-axis ($R_x(\delta)$), 
is performed to bring the points symmetrically
about the $z$ axis to finally 
arrive at the configuration of points 
for the canonical state $\psi_c$. The angles of
rotations are given by:
\begin{eqnarray}
 \beta &=& \tan^{-1} \frac{\cos \phi_1 \tan \theta_1-\cos \phi_2 \tan \theta_2}
{\sin \phi_1 \tan \theta_1-\sin \phi_2 \tan \theta_2}  \nonumber \\
 \gamma &=& \cos^{-1} \pm \sqrt{\frac{\cos^2 \theta_1-\cos^2 \theta_2}
{\sin^2 \theta_2 \sin^2 (\beta+\phi_2) -\sin^2 \theta_1 \sin^2 (\beta+\phi_1) }} \nonumber \\
 \delta &=& \tan^{-1} \frac{\cos \gamma (\sin \theta_1 \sin (\beta+\phi_1)
+\sin \theta_2 \sin (\beta+\phi_2)}
{\cos \theta_1 + \cos \theta_2} \nonumber \\
\end{eqnarray}
Corresponding operations in the Hilbert space leads us to 
the canonical state, starting from the general spin-1 state
(or vice versa)
\begin{equation}
\vert \psi_c \rangle = U_x(\delta)U_z(\gamma)U_y(\beta) \vert \psi_g \rangle.
\end{equation}
Thus, the single parameter canonical state 
($\vert \psi_c \rangle_\alpha$) alongwith 
the rotations by angles $\beta$, $\gamma$,
and $\delta$ can be used to reach any state 
in the three-dimensional Hilbert space.
\subsection{Invariant Magnetization}
\label{magnetization-vector} 
When a magnetic moment is associated
with a qutrit,  we can associate magnetization with
the corresponding angular momentum operators.
For a general pure state of the qutrit $\rho =
\vert \psi \rangle \langle \psi \vert$ the
magnetization vector in terms of the expectation
values of $\Sigma_i$ can be  written as:
\begin{equation}
 \overrightarrow{M} = \langle \Sigma_1 \rangle \hat{x}
+ \langle \Sigma_2 \rangle \hat{y} 
+ \langle \Sigma_3 \rangle \hat{z}.
\end{equation}
For the canonical state
$\vert \psi_c \rangle$ the expectation 
values  are
\begin{equation}
\langle \Sigma_1 \rangle = 0, \quad \langle
\Sigma_2 \rangle = 0, \quad
\langle \Sigma_3 \rangle = -\cos(2\alpha).
\end{equation}
and the  magnitude of the magnetization vector is:
\begin{equation}
\vert \overrightarrow{M} \vert = 
\vert \cos(2\alpha) \vert  \label{mag}.
\end{equation}
For the eigen states of $\Sigma_3$ the values are 
\begin{equation}
{\rm For }\,\, \alpha=0, \, \frac{\pi}{2}\,  \vert
\overrightarrow{M} \vert = 1, \quad {\rm For}\,\,
\alpha=\frac{\pi}{4},\, \vert \overrightarrow{M}
\vert = 0.
\end{equation}
The action of $SO(3)$  causes rotation of spin as
a whole without affecting the magnitude of the
magnetization vector.  Thus the amount of
magnetization of a qutrit in an arbitrary state
$\vert \psi \rangle$ remains unaltered under
$SO(3)$.

The magnetization finds a natural representation
on the Majorana sphere where the magnetization
vector or a
single qutrit can be related to the bisector of
the angle enclosed between the pair of points
representing the state on the Majorana sphere.
As per Fig.~\ref{canonical-rep}, the length
$l_b$
of the perpendicular bisector $OO'$ 
is 
\begin{equation}
l_b=z_c=\cos{\frac{\eta}{2}}
=\frac{\sin \alpha - \cos \alpha}{\sin
\alpha + \cos \alpha}.
\end{equation}
The magnetization $\vec{M}$, the canonical state parameter
$\alpha$ and the length of the bisector $OO'$
are related to each other as  
\begin{equation}
 \vert \overrightarrow{M} \vert = \vert \langle
\psi_c \vert \Sigma_3 \vert \psi_c \rangle \vert
 = \vert-\cos(2 \alpha) \vert
 = \frac{2 \vert l_b \vert}{(l_b^2 + 1)} 
\end{equation}
Thus, in the context of Majorana representation,
$\vert \overrightarrow{M} \vert$ can be obtained
from length $l_b$ of the bisector $OO'$
(Fig.~\ref{canonical-rep}).  
Under the action of  $SO(3)$ the quantities $l_b$
and $\frac{2 \vert l_b \vert}{(l_b^2 + 1)}$ are
invariant.

\begin{figure}
\centering
\includegraphics[scale=1]{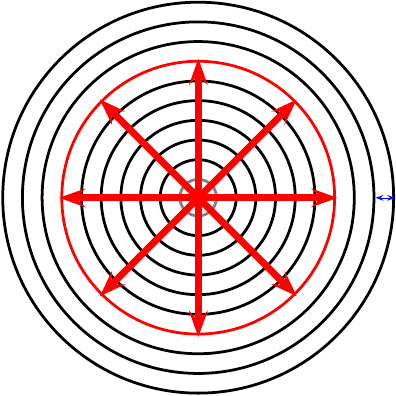}
\caption{Cross-sectional representation of the
Majorana sphere.  Circles of different radii
correspond to surfaces of constant magnetization
of a single qutrit. Red arrows pointing towards
different directions are a few of the infinitely
many magnetization vectors for a given value of
the magnitude of  magnetization (shown in
red).}\label{magvec}
\end{figure}

The magnitude of the magnetization vector $\vert
\overrightarrow{M} \vert$ in a pure ensemble of a
single qutrit can assume values in the range
$[0,1]$. On the contrary, the pure ensemble of a qubit
always possesses unit magnitude of the
magnetization vector associated with it.  The
geometrical picture of the single qutrit
magnetization vector is provided by the Majorana
representation.
The value $\vert \overrightarrow{M} \vert$ depends
upon the length of the bisector $OO'$ and lies along
the $z$-axis and is rotationally
invariant.  Thus corresponding to a given value of
the length of the bisector, one can assume
concentric spheres with continuously varying
radii, whose surfaces are the surfaces of constant
magnetization. The radii of these spheres are
equal to $\vert \overrightarrow{M} \vert$, that
vary in the range [0,1]. Circular cross-section of
the Majorana sphere depicting the magnetization
vector is shown in Fig.~\ref{magvec} where an
example with $\vert \overrightarrow{M} \vert =
\frac{1}{\sqrt{2}}$ is shown as a red circle
inside the cross-section of a unit sphere.
Depending on whether a specific direction can be
assigned to the qutrit state or not, these states
can be classified as `pointing' or `non-pointing'.
A single qutrit state that has a direction of
orientation and also possesses a non-zero
magnetization is termed a pointing state, while
the states with zero magnetization content and
without any directional orientation are termed
non-pointing states.  In the  Majorana
representation, non-pointing states consist of
diametrically opposite points on the unit sphere,
such that the bisector of the angle enclosed at
the center of the sphere vanishes. On the other
hand, pointing states have non-zero magnetization.
Under $SO(3)$ transformations, pointing states
remain always pointing while the non-pointing
states remain non-pointing.  For example the
states  $\vert +1 \rangle$($\vert -1 \rangle$),
state points along $+z$($-z$) direction. On the
other hand $\vert 0 \rangle$ has diametrically
opposite points in the Majorana representation and
is a non-pointing state.

In general, an eigenstate of $\Sigma_i$ with
eigenvalue $+1$ points along the $+i$ direction
and an eigenstate of $\Sigma_i$ with eigenvalue
$-1$ points in the $-i$ direction. States with
zero eigenvalue can be supposed to have a zero
magnetization and thus do not have any preferred
direction of orientation.

The canonical state of a single-qutrit undergoes 
a rotation under the effect of $SO(3)$ transformations,
in such a way that the magnitude of the magnetization
vector remains unchanged. However $SU(3)$ transformations
change both the magnitude as well as direction of the 
single-qutrit magnetization vector. For instance, 
a unitary transformation, $e^{-\iota \theta
\Lambda_5}$,
transforms  the coefficients of
$\vert +1 \rangle$ and $\vert -1 \rangle$ of 
the one-parameter family  of states $\vert \psi_c
\rangle$ in such a way that the angle $\alpha$.
Therefore, this type of operation can be used to
alter the magnetization magnitude.
On the Majorana sphere, this operation corresponds to 
opening or closing of angle $\alpha$ between the two 
points representing the canonical state of the
qutrit as depicted in Fig.~\ref{canonical-rep}.

The magnetization for the general state  $\vert
\psi_g \rangle$ defined in 
Eq.~(\ref{genstate}) can be easily computed and
turns out to be:
\begin{eqnarray}
\langle \Sigma_1 \rangle &=& 
\Gamma (\sin \theta_1 \cos \phi_1 + \sin \theta_2 \cos \phi_2) \nonumber \\
\langle \Sigma_2 \rangle &=& 
\Gamma (\sin \theta_1 \sin \phi_1 + \sin \theta_2 \sin \phi_2) \nonumber \\
\langle \Sigma_3 \rangle &=& 
\Gamma (\cos \theta_1 + \cos \theta_2) 
\end{eqnarray}
The Majorana representation of the general single-qutrit 
state is given by two points:
$P_1(\sin \theta_1 \cos \phi_1, \sin \theta_1 \sin \phi_1, 
\cos \theta_1)$ and $P_2(\sin \theta_2 \cos \phi_2, 
\sin \theta_2 \sin \phi_2, \cos \theta_2)$.
Point $O'(x',y',z')$ bisects the chord $P_1P_2$, which when 
connected to the center of the sphere $O$, provides 
the direction of the single-qutrit magnetization
vector:
\begin{equation}
 \overrightarrow{M} =
 \langle \Sigma_1 \rangle \hat{x} + \langle \Sigma_2 
 \rangle \hat{y} + \langle \Sigma_3 \rangle \hat{z}
  = \frac{2}{l_b^2+1} \overrightarrow{OO'}.
\end{equation}
Thus a physical picture of the single-qutrit 
magnetization vector is realized  with the help of 
Majorana representation of a qutrit. Further, one 
can observe the evolution of single-qutrit magnetization 
vector under various quantum operations.
\section{Quantum computing with a single-qutrit}
\label{computing}
\subsection{NMR qutrit}
\label{qutrit}
\begin{figure}[h!]
\includegraphics{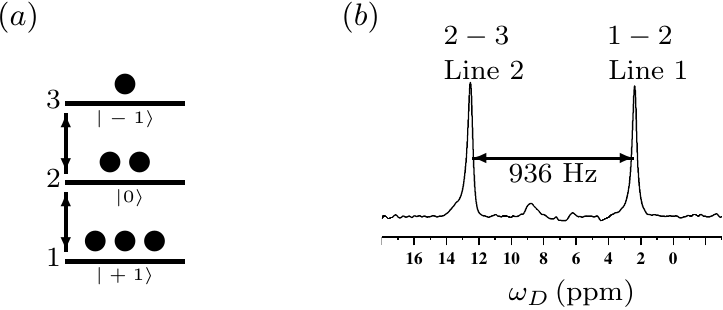}
\caption{
(a) Single-qutrit energy level diagram 
depicting the thermal equilibrium population distribution
of the eigen vectors labeled as $\{ 1,2,3\}$. 
(b) Deuterium NMR spectrum of oriented deuterated chloroform 
molecule at 277 K. The spectral
lines are labeled as Line~1 and Line~2. At 277 K, the anisotropic 
liquid crystalline environment is shown to 
exhibit an effective quadrupolar splitting of 936
Hz.~\label{system}}
\end{figure}

A natural realization of  a qutrit in NMR is a spin-1
nucleus, wherein the three eigenstates of the operator
$\Sigma_3$ namely $\vert +1 \rangle$, $\vert 0 \rangle$ and
$\vert -1 \rangle$, split in  energy in the presence of a
magnetic field as shown in Fig.~\ref{system}(a) (the states
are labeled as $\{1,2,3\}$ for convenience).
The NMR Hamiltonian of a spin-1 nucleus
in the presence of a 
static magnetic field, $B_0=-\omega_0/\gamma$,
is given
by~\cite{levitt-book-2008,slichter-book-1996}
(we have used the standard NMR notation in this
section and the $I$ operators in this section are
the same as the $\Sigma$ operators of the
previous section):
\begin{equation}
H = - \omega_0 I_z + \frac{e Q
V_{zz}}{4I(2I-1)}(3I_z^2-I^2) \label{ham}
\end{equation}
where $\omega_0$ is the Larmor precession frequency,
$\gamma$ is the gyromagnetic ratio, $Q$ is the quadrupolar
moment and $V_{zz}$ is the average value of the uniaxial
electric field gradient component over the molecular motion.
The second term in the R.H.S. of Eq.~(\ref{ham}) is the
quadrupolar coupling term, whose average vanishes in an
isotropic medium as provided by liquid state NMR. Only the
Zeeman interaction term survives, leading to three equally
separated energy levels and an overlapped pair of spectral
lines  which cannot be addressed separately in frequency
space.  The quadrupolar coupling is retained in an
anisotropic environment such as that provided by a liquid
crystalline medium.  The anisotropic molecular orientation
with respect to the magnetic field gives rise to a
quadrupolar coupling term in the Hamiltonian    which is now
given by
\begin{equation}
H = - \omega_0 I_z
+ \kappa (3 I_z^2 - I^2) 
\end{equation}
where $\kappa = e^2 q Q S / 4$ is the effective value of
the quadrupolar coupling, $e q =V_{zz}$ denotes the field
gradient parameter, and $S$ is the order parameter of the
liquid crystal.  The effective quadrupolar coupling value
depends upon the order parameter of the liquid crystal, and
is responsible for the splitting between the previously
degenerate energy levels which now gives rise to two
non-degenerate spectral lines.

The spin-1 system used in this study is the deuterium spin
in a deuterated chloroform molecule oriented in a lyotropic
liquid crystal~\cite{dogra-pla-2014}.  The lyotropic liquid
crystal is composed of $25.6 \%$ of Potassium Laurate,
$68.16 \%$ of H$_2$O and $6.24 \%$ of
Decanol~\cite{yu-prl-1980}, and 50 $\mu l$ of Chloroform-D
was added to 500 $\mu l$ of the liquid crystal during sample
preparation.  A  temperature of 277 K was found to be
optimal for performing the experiments (at which $T_1$ and
$T_2$ relaxation times are approximately $170$ ms and $50$
ms respectively). The relative separation of the single
quantum transitions is  $936$ Hz on a $600$ MHz NMR
spectrometer (as shown in Fig.~\ref{system}(b)).
All the experiments were performed on a 600 MHz Avance III
NMR spectrometer equipped with a QXI probe, with the
deuterium nucleus resonating at 91.108 MHz.  All the
transition-selective rf pulses used in this work are 
'Gaussian' shaped pulses of $4$ ms
duration, while the non-selective rf
pulses are 'Sinc' shaped pulses with a duration of $0.5$ ms. 
\subsection{Initialization and tomography of a single qutrit}
\label{initial}
\begin{figure}[h!]
\includegraphics{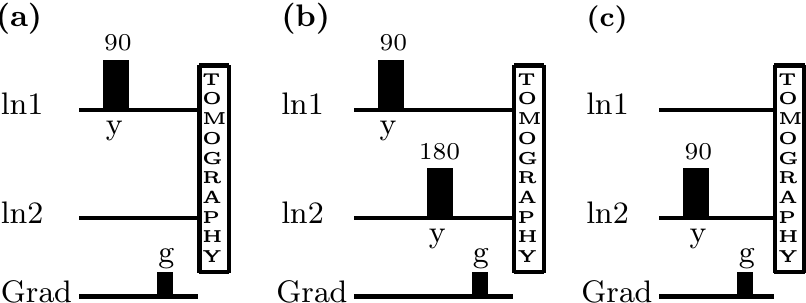}
\caption{Pseudopure state preparation scheme, 
resulting into states : (a) $\vert -1 \rangle$, (b) 
$\vert 0 \rangle$ and (c) $\vert +1 \rangle$. Channels
ln1 and ln2 correspond to transitions
1-2 and 2-3 respectively. The gradient pulse of strength $g$
is shown in the third channel. \label{pps}}
\end{figure}

For the purpose of NMR quantum computation, one 
needs to initialize the quantum system in a 
pseudopure state which is obtained by
having a different population in one of the energy levels over 
a uniform background in an ensemble of $\sim10^{18}$ spins.
A qutrit has three 
possible pseudopure states : $\vert +1 \rangle, \,
 \vert 0 \rangle, \, \vert -1 \rangle$, which are
obtained from the thermal equilibrium state using 
transition-selective pulses followed by a gradient 
pulse that dephases the 
coherences~\cite{das-ijqi-2003,dogra-pla-2014}. 
NMR pulse sequences for 
the preparation of pseudo-pure states 
$\vert -1 \rangle, \,
 \vert 0 \rangle, \, \vert +1 \rangle$ are shown in 
Fig.~\ref{pps}(a),(b),(c) respectively. 

\begin{figure}[h!]
 \centering
\includegraphics[scale=1]{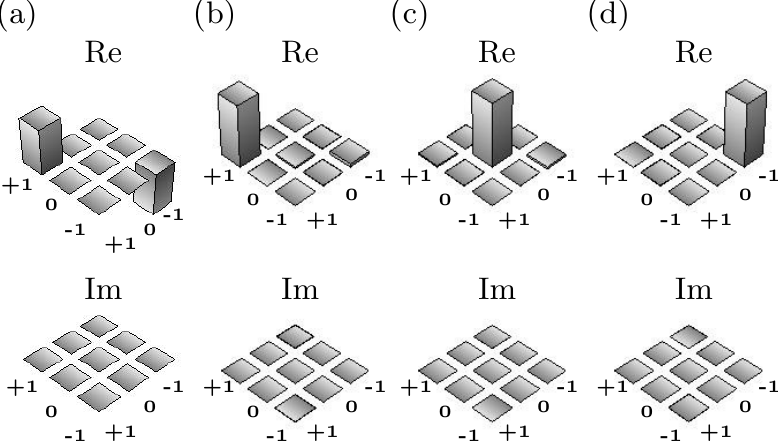}
\caption{Real (Re) and imaginary (Im) parts of the experimentally 
constructed single-qutrit density matrices are shown for 
(1) thermal equilibrium state, and the pseudo-pure states:
(b) $\vert +1 \rangle$, (c) $\vert 0 \rangle$,
and (d) $\vert -1 \rangle$.  These 
pseudo-pure states are created using 
pulse sequences given in Fig.~\ref{pps}. 
The pseudo-pure states $\vert +1 \rangle$, $\vert 0 \rangle$,
and $\vert -1 \rangle$ are found to have fidelities of 0.99, 0.97 
and 0.99 respectively.
\label{pps-tomo}}
\end{figure}

The final readout of the experimental state is
performed
by quantum state tomography. Generators of $SU(3)$
operators ($\Lambda$ matrices) alongwith a $3 \times
3$ identity matrix form a complete set of orthonormal
bases in the operator space of spin-1.
The state of the spin-1 in general is
$\rho = \frac{1}{2}\sum_{i=1}^{8} c_{i}
\Lambda_{i}$ where $c_{i}$s are real numbers.
State reconstruction requires all eight 
expectation values i.e. $c_i$s to be determined, which
requires a minimum of four operations which are given
below alongwith their corresponding NMR pulse 
sequences~\cite{dogra-pla-2014}:
\begin{itemize}
\item Identity : No operation 
\item $U_{\Lambda_1}(\frac{3\pi}{2})$ :
$I_x^{1-2}(-\pi)$ 
\item kill
coherences, $U_{\Lambda_1}(\frac{\pi}{4})$ : $Grad_z$,
$I_y^{1-2}(\frac{\pi}{2})$ 
\item kill coherences, $U_{\Lambda_7}(\frac{\pi}{4})$
: $Grad_z$, $I_y^{2-3}(\frac{\pi}{2})$ 
\end{itemize}
The NMR spectrum of a qutrit contains of two lines
corresponding to the single quantum coherences $\rho_{12}$ and 
$\rho_{23}$. The first experiment in the tomography
protocol outlined above gives the values of
($c_1, c_2, c_6, c_7$), the second experiment interchanges
the intensities of single- and double-quantum terms 
providing the values of ($c_4, c_5$) and the last two
experiments give the diagonal elements and hence
the values of ($c_3, c_8$). Calculating all these 
values from the intensities of spectral lines, the 
experimental density matrix ($\rho_e$) can be constructed.
The measure of overlap used between the
theoretically expected and experimentally constructed
density matrices is termed the fidelity $F$:
\begin{equation}
  F = \frac{Tr(\rho \dag
\rho_e)}{\sqrt{Tr(\rho \dag \rho)}\sqrt{Tr(\rho_e
\dag \rho_e)}}. \label{fidelity}
\end{equation}
The experimentally tomographed single-qutrit 
states are shown in Fig.~\ref{pps-tomo}.
Pseudopure states $\vert +1 \rangle$, $\vert 0 \rangle$
and $\vert -1 \rangle$ are obtained with the fidelities 
0.99, 0.97 and 0.99 respectively.
\subsection{Experimental implementation of $\Lambda$
matrices}
\label{exptl}
\begin{figure}[h!]
\includegraphics{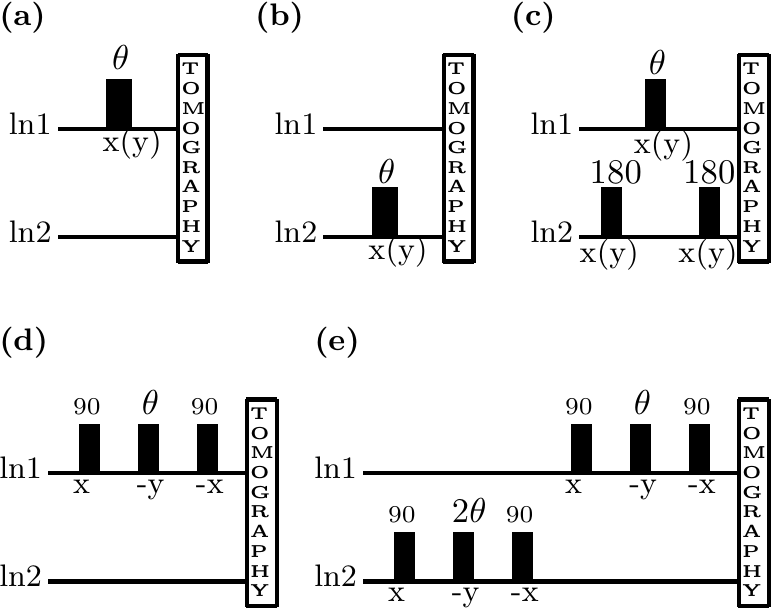}
\caption{NMR pulse sequence for the implementation of
(a) $U_{\Lambda_1}(U_{\Lambda_2})$ giving rotation
$\frac{\theta}{2}$, using transition selective pulses. 
$\theta=\pi$ swaps the populations between energy 
levels 1 and 2 of a single spin 1, 
(b) $U_{\Lambda_6}(U_{\Lambda_7})$
giving rotation $\frac{\theta}{2}$, using transition 
selective pulses. $\theta=\pi$ swaps the populations 
between energy levels 2 and 3, 
(c) $U_{\Lambda_4}(U_{\Lambda_5})$ 
giving rotation $\frac{\theta}{2}$, using a single spin 
selective pulse. $\theta=\pi$ swaps the populations between 
energy levels 1 and 3. Pulse sequences for the 
implementation of (d) $U_{\Lambda_3}$ and 
(e) $U_{\Lambda_8}$ matrices giving rise 
to the relative phase shifts. All the black rectangular 
pulses are the shaped pulses with angle of rotation 
mentioned at the top and the axis of rotation is 
mentioned at the bottom. ln1 and ln2 stand for the 
transitions between levels 1-2 and 2-3 respectively.
\label{lambda-pp}}
\end{figure}

Each of the two single-quantum transitions of an
NMR qutrit
belong to a two-level subspace.
The transition-selective pulses in NMR are
governed by transition-selective operators as described 
earlier (Eq.~(\ref{n1-n3})).
The experimental implementations of the
$\Lambda$ matrices $U_{\Lambda_1}$ and $U_{\Lambda_2}$ 
require a selective excitation of the transition between
energy levels $1$, $2$, while the
implementations for $U_{\Lambda_6}$ and
$U_{\Lambda_7}$ 
are achieved by a transition-selective radio frequency pulse
between energy levels $2$, $3$. 
The implementations for
$U_{\Lambda_4}$ and
$U_{\Lambda_5}$ 
require a double-quantum excitation which is 
achieved using a set of three transition-selective 
pulses~\cite{dorai-jmr-1995}.
$U_{\Lambda_3}$ and $U_{\Lambda_8}$ correspond to $z$-rotations
and can be implemented using $z$-cascade 
pulses. 
The NMR pulse 
sequences corresponding to the $U_{\Lambda_i}(\theta)$ implementations
are shown in Fig.~\ref{lambda-pp}.
Implementation of unitary transformations
corresponding to $\Lambda$ matrices on 
single-qutrit magnetization vector, leads to change in 
both its magnitude as well as direction and can be
visualized on the Majorana sphere as described in
the Section~\ref{magnetization-vector}.
\subsection{Quantum Gates}
\label{qgates}
Similar to qubits, quantum computation with qutrits uses
single-qutrit and two-qutrit operators
as quantum gates. We describe below the
implementation of some basic
gates on an NMR qutrit.
\subsubsection{Chrestenson gate}
\begin{figure}[h!]
\centering
\includegraphics{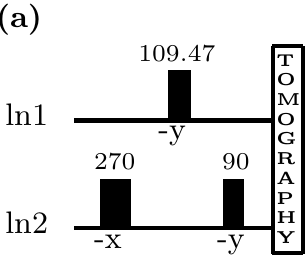}
\caption{Pulse sequence for Chrestenson gate
 implementation on qutrit pseudopure states. 
The two 
channels correspond to two transitions of a
 single qutrit. Pulse
 angles and axes are  mentioned alongwith all 
the pulses.\label{chrestenson-pp}}
\end{figure}
The Chrestenson gate is the qutrit analog of the
single-qubit Hadamard gate.  It is a single-qutrit
gate that creates uniform superpositions of the
energy levels alongwith relative phases between
the basis vectors.  Chrestenson gate for
single-qutrit is written as:
\begin{eqnarray} 
Ch = \frac{1}{\sqrt{3}} \left( \begin{array}{ccc}
1 & 1 & 1 
\\
 1 & e^{\frac{2\iota\pi}{3}} & e^{\frac{4\iota\pi}{3}}  
\\
 1 & e^{\frac{4\iota\pi}{3}} & e^{\frac{2\iota\pi}{3}} 
\end{array}  \right)
\end{eqnarray} 
Considering the action of Chrestenson
gate on the non-pointing state $\vert 0 \rangle $
leads to the state
$\frac{1}{\sqrt{3}} \left( \vert +1 \rangle + e^{\frac{2 \pi
\iota}{3}}\vert 0 \rangle + e^{\frac{4 \pi \iota}{3}}
\vert -1 \rangle \right)$, which is a pointing state.
Chrestenson gate can not be obtained solely by 
$SO(3)$ transformations. Therefore, the decomposition of
this gate 
requires more general$SU(3)$ operators which in turn require
transition-selective pulses for NMR experimental 
implementation.
We proposed a non-trivial sequence of rf pulses for the 
Chrestenson gate implementation in NMR (details
given in Ref.~\cite{dogra-pla-2014}).
 \begin{figure}[h!]
 \centering
\includegraphics[scale=1]{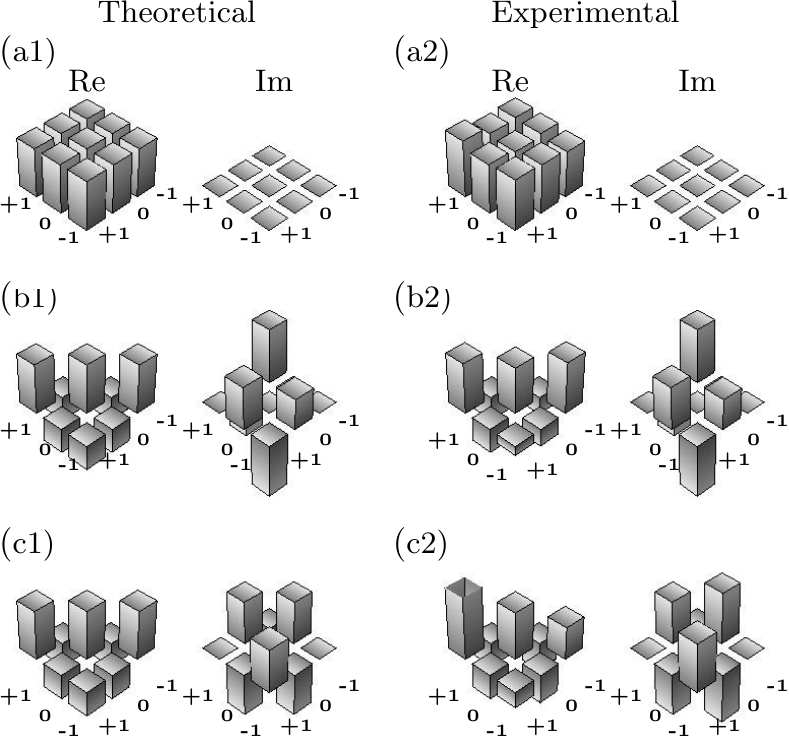}
\caption{Real and imaginary parts of the 
theoretically expected ($a_1, b_1, c_1$) and 
experimentally obtained ($a_2, b_2, c_2$) tomographs 
of the resultant single-qutrit state after
Chrestenson gate implementation on different initial
states. The initial states corresponding to
different parts are: ($a_1,a_2$) 
$\vert +1 \rangle$, ($b_1,b_2$) $\vert 0 \rangle$,
and ($c_1,c_2$) $\vert -1 \rangle$.
\label{chrestenson-tomo}}
\end{figure}

\begin{figure}
\includegraphics{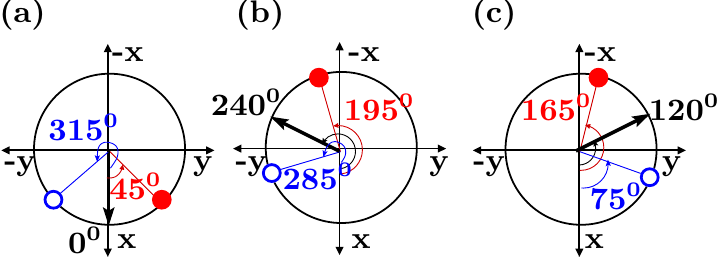}
\caption{
Points on the  $z=0$ cross-section of the
Majorana sphere representing the state of the qutrit after
the implementation of a Chrestenson gate.  The pair of points
solid circle (red) and the empty circle (blue) show the
points after application of the gate on the states $|+1
\rangle$, $|0 \rangle$, and $|-1 \rangle$ as depicted in
panels (a), (b), and (c) respectively. The thick black arrow 
represents the direction of the magnetization vector (of
magnitude $\frac{2\sqrt{2}}{3}$) in all the cases.
\label{ch-maj}}
\end{figure}

This pulse sequence is applied on all three
qutrit bases states and the final state is completely
tomographed. Final state density matrices are 
shown in Fig.~\ref{chrestenson-tomo}. As seen from the 
Fig.~\ref{chrestenson-tomo}, theoretically expected 
and experimentally obtained single-qutrit states 
match well after the Chrestenson gate
implementation. The exact overlap between the 
theoretically expected and experimentally obtained 
states (Eq.~(\ref{fidelity})) is found out to be
0.99, 0.98, 0.98 corresponding to initial states 
$\vert +1 \rangle$, $\vert 0 \rangle$, 
$\vert -1 \rangle$ respectively.
Visualizing on the Majorana sphere,
Chrestenson gate implementation on all 
three basis states of a single-qutrit, brings the 
magnetization vector to the plane $z=0$. The 
magnetization vector of these respective final 
states lie symmetrically in the plane $z=0$ at 
angles of $2\pi/3$ with respect to each other.
This is shown in Fig.~\ref{ch-maj}, where the pair of points
lie on a cross-section of the Majorana sphere defined by $z=0$.
Figs.~\ref{ch-maj}$(a),(b),(c)$ correspond to the Chrestenson
gate implementation on the bases states $|+1 \rangle$, $|0
\rangle$, and $|-1 \rangle$, respectively.  The bisectors of
the angles enclosed by the pair of Majorana points which
represents the spin magnetization are shown as thick black arrows
in each case. The magnitude of the magnetization vector in
each case is $|\overrightarrow{M}|=\frac{2\sqrt{2}}{3}$,
while the directions are different.

\subsubsection{SWAP gates}
\begin{figure}[h!]
\centering
\includegraphics{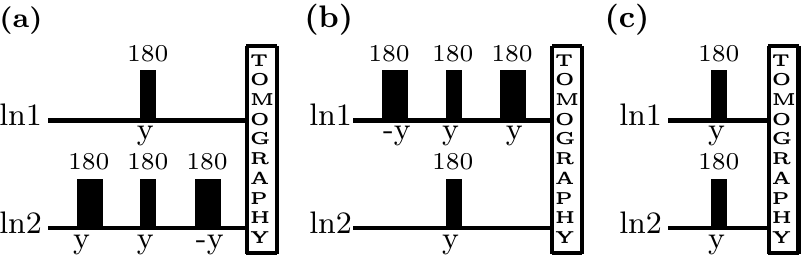}
\caption{Pulse sequence for implementation 
of swap operations. (a) $\rm{SWAP}_{12}$, (b)
$\rm{SWAP}_{23}$ and 
(c) $\rm{SWAP}_{13}$. Transition selective pulses are shown 
by thick rectangles while thin rectangles are 
the non-selective pulses.\label{swap-pp}}
\end{figure}

A qutrit is a three-level system, therefore there are
three possible swap operations between levels 1-2, 2-3
and 1-3. This is achieved by using transition-selective
pulses. Exact pulse sequences for NMR implementation of 
$\textrm{SWAP}_{12}$, $\textrm{SWAP}_{23}$ and 
$\textrm{SWAP}_{13}$ are shown 
in Fig~\ref{swap-pp}. However for $\theta=\pi$,
SWAP gates can effectively be implemented by the
unitary operators $U_{\Lambda_j}$.
The explicit matrices for
$\textrm{swap}_{12}$, $\textrm{swap}_{23}$ and $\textrm{swap}_{13}$ are 
written as 
\begin{equation}
 \left(
\begin{array}{lll}
 0 & 1 & 0 \\
 1 & 0 & 0 \\
 0 & 0 & 1
\end{array}
\right),
\quad
 \left(
\begin{array}{lll}
 1 & 0 & 0 \\
 0 & 0 & 1 \\
 0 & 1 & 0
\end{array}
\right),
\quad
 \left(
\begin{array}{lll}
 0 & 0 & 1 \\
 0 & 1 & 0 \\
 1 & 0 & 0
\end{array}
\right)
\end{equation}

\begin{figure}[h!]
 \centering
\includegraphics[scale=1]{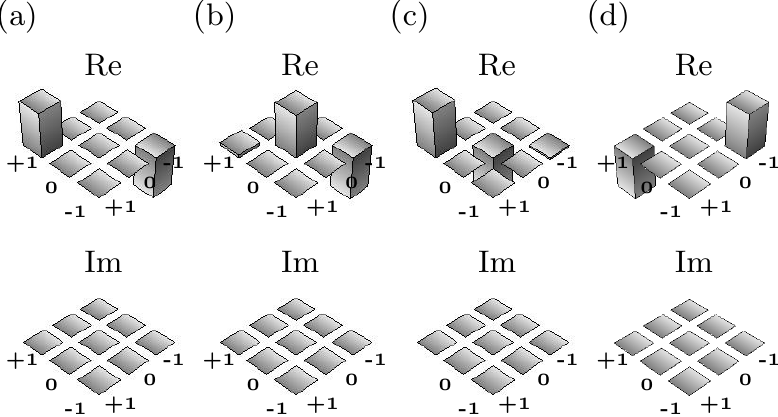}
\caption{Real and imaginary parts of the 
(a) Initial state (thermal equilibrium) 
and the states resulting after 
the implementation of (b) $\textrm{SWAP}_{12}$,
(c) $\textrm{SWAP}_{23}$,
and (d) $\textrm{SWAP}_{13}$ operators, 
having state fidelities of 0.99, 0.99 and 
0.98 respectively.
\label{swap-tomo}}
\end{figure}

\begin{figure}[h!]
\includegraphics{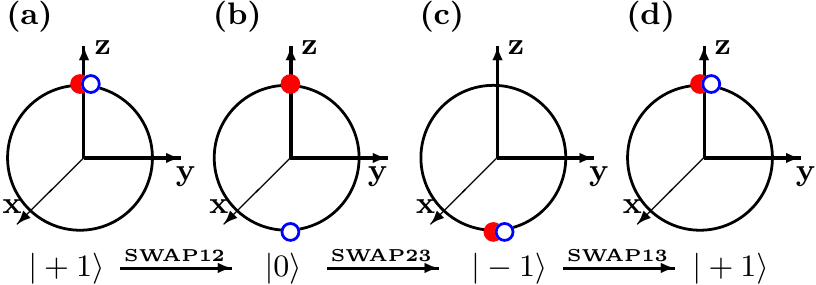}
\caption{
The sequential action of $\textrm{SWAP}_{12}$,
$\textrm{SWAP}_{23}$, and $\textrm{SWAP}_{13}$ gates on the
state $|+1\rangle$ shown on the Majorana sphere.  Majorana
representative points of the initial state are shown in (a)
as a solid circle (red) and an empty circle (blue). The
panels (b,c,d) show the two points after the action of
$\textrm{SWAP}$ operations carried out sequentially.
\label{swap-maj}}
\end{figure}

All three SWAP gates are implemented experimentally 
on the thermal equilibrium state. Complete 
quantum state tomography of all states was
performed.
The fidelities corresponding to
$\textrm{SWAP}_{12}$, $\textrm{SWAP}_{23}$, 
and $\textrm{SWAP}_{13}$ implementations
are 0.99, 0.99, 0.99 respectively. The tomographs of
the initial and final states are shown in Fig.~\ref{swap-tomo}.
$\textrm{SWAP}$ gates are implemented on single-qutrit states, whose
Majorana representation consists of two points on the unit sphere.
Under $\textrm{SWAP}_{12}$ gate implementation one of the points 
(say the first point) gets flipped and occupies the diametrically 
opposite position on the Majorana sphere; $\textrm{SWAP}_{23}$ 
implementation leads to the flipping of second point in 
Majorana representation while under $\textrm{SWAP}_{13}$ operation, 
both the Majorana points get flipped.  The
actions of various $\textrm{SWAP}$ gates are shown 
on the Majorana sphere in Fig.~\ref{swap-maj}, where we
have taken the initial state to be $|+1\rangle$ and
subjected it to consecutive $\textrm{SWAP}$ gates.
\subsubsection{Phase gates \label{phase-gate}}
\begin{figure}[h!]
\includegraphics{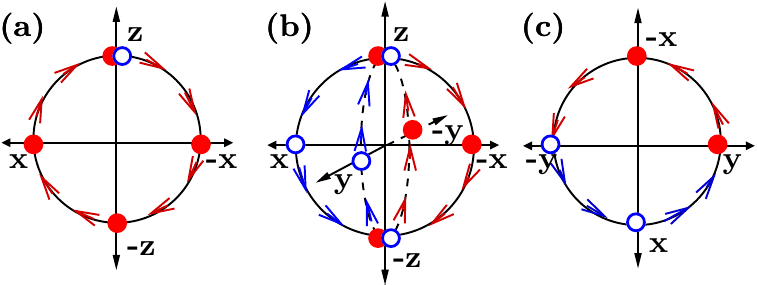}
\caption{
The trajectories of the Majorana points under the
transformations generated by $\Lambda_2$, $\Lambda_5$, and
$\Lambda_3$ matrices. Panel (a) shows the evolution of the
state $|+1\rangle$ under $e^{\iota \frac{\theta}{2}
\Lambda_2}$.  As $\theta$ varies from $0$ to $2\pi$, the
unfilled blue point does not move at all, while the 
solid red point moves along a circle in the $x-z$ plane.  
Panel (b) shows the evolution of the state
$|+1\rangle$ under $e^{\iota \frac{\theta}{2} \Lambda_5}$. In
this case both points (solid red and unfilled blue) move in
opposite directions simultaneously along a cirlce in the
$x-z$ plane.  They meet at the south pole when $\theta=180^0$
and then come back via the $z-y$ plane. Panel (c) shows  the
trajectory followed  by the state
$\frac{1}{\sqrt{2}}(|+1\rangle+|-1\rangle)$ under $e^{\iota
\theta \Lambda_3}$.  In this case both the points move in
the $x-y$ plane (starting from $y=\pm1$ in the same direction
and by same amounts as $\theta$ varies. 
\label{lambda-maj}}
\end{figure}

\begin{table}[h!]
\caption{Phase differences between the two single 
quantum transitions obtained corresponding to different
values of $\theta$ under $\Lambda_3$ and $\Lambda_8$ 
implementations. Columns 3 and 4 contain the theoretically
expected and experimentally obtained values of the 
phase difference obtained in $\Lambda_3$ implementations,
and columns 5,6 contain the theoretically
expected and experimentally obtained values of the 
phase difference obtained as a result of
$\Lambda_8$ implementations. The angles are given
in degrees. \label{table}}
\begin{center}
\renewcommand{\arraystretch}{1.2}
\begin{tabular}{|l|l|l|l|l|l|}
\hline
\multicolumn{2}{|c|}{} & \multicolumn{2}{c|}{$U_{\Lambda_3}$}
&\multicolumn{2}{c|}{$U_{\Lambda_8}$} \\
\hline
& $\theta~~~$  & $\theta_{th}~~~$   &
 $\theta_{exp}~~~$   & $\theta_{th}~~~$  &
$\theta_{exp}$ \\
 \hline
 \hline
 ~1~~ & ~0~   & ~0~    & ~0~     & ~0~
& ~0 \\
 ~2 & ~30~  & ~45~   & ~39.95~ & ~30$\sqrt{3}$~  &
~47~ \\
 ~3 & ~45~  & ~67.5~ & ~62.4~  & ~45$\sqrt{3}$~  &
~70~ \\
 ~4 & ~60~  & ~90~   & ~84.2~  & ~60$\sqrt{3}$~  &
~110~ \\
 ~5 & ~90~  &  ~135~ & ~139.05~& ~90$\sqrt{3}$~  &
~168~ \\
 ~6 & ~120~ & ~180~  & ~203.3~ & ~120$\sqrt{3}$~ &
~220~ \\
 \hline 
\end{tabular} 
\end{center}
\end{table}
Phase shift gates in qutrits introduce
relative phases between the base vectors of the same qutrit.
These gates are primarily $z-$rotations and are
obtained
by  $U_{\Lambda_3}(\theta)$ and $U_{\Lambda_8}(\theta)$
implementations, with their explicit matrix forms
being:
\begin{equation}
 U_{\Lambda_3} =\left(
\begin{array}{lll}
 e^{\iota \theta} & 0 & 0 \\
 0 &  e^{-\iota \theta} & 0 \\
 0 & 0 & 1
\end{array}
\right),
\quad
U_{\Lambda_8} =\left(
\begin{array}{lll}
 1 & 0 & 0 \\
 0 & 1 & 0 \\
 0 & 0 &  e^{\iota \sqrt{3}\theta}
\end{array}
\right)
\end{equation}
These gates are implemented with various values of
$\theta$. The initial
state ($\rho_i^{exp}$) is created by a
non-selective $90^{\circ}$ pulse on thermal equilibrium
state, so that one begins with in-phase values of
the single quantum coherences. Action of
$U_{\Lambda_3}(\theta)$ builds a relative phase
difference of $\frac{3\theta}{2}$ between the two
single quantum coherences. This relative phase is
the phase difference between the spectral lines in
a single-qutrit NMR spectrum obtained by
$U_{\Lambda_3}(\theta)$ implementation on
$\rho_i^{exp}$.  The second column of the
Table~\ref{table} shows the values of angles
implemented ($\theta$) as given in
Fig~\ref{lambda-pp}(d), columns 3 and 4 contain
the values of the theoretically expected
($\theta_{th}$) and experimentally obtained
($\theta_{exp}$) values of the phase difference
respectively.  A $U_{\Lambda_8}(\theta)$ builds a
relative phase difference of $\sqrt{3}\theta$
between the two single-quantum coherences.
Relative phase in the spectral lines of the first
order single-qutrit spectrum under
$U_{\Lambda_8}(\theta)$ implementation
($\theta_{exp}$) is shown in column 6 of the
Table~\ref{table}, column 5 contains the
theoretically expected values of phase difference
($\theta_{th}$) between the two single quantum
coherences.
Fig.~\ref{lambda-maj} displays the action
of unitary transformations generated by various $\Lambda$
matrices on the Majorana sphere. In Fig.~\ref{lambda-maj}(a)
and (b) we consider the transformations generated by
$\Lambda_2$ and $\Lambda_5$ on the state $\vert + \rangle$.
In  Fig.~\ref{lambda-maj}(c) we show the  trajectory
corresonding to the transformations generated by
$U_{\Lambda_3}$ which represents the phase gate.  The initial
state in this case is taken to be
$\frac{1}{\sqrt{2}}(|+1\rangle+|-1\rangle)$. The
corresponding NMR pulse sequences for these trajectories are
given in Fig.~\ref{lambda-pp}.
\section{Concluding Remarks}
\label{conclusions}
A geometrical representation of a qutrit is
described in this work, wherein qutrit states are
represented by two points on a unit sphere as per
the Majorana representation. A parameterization of
single-qutrit states was obtained to generate
arbitrary states from a one-parameter family of
canonical states via the action of $SO(3)$
transformations. The spin-1 magnetization vector
was represented on the Majorana sphere and states
were identified as `pointing' or `non-pointing'
depending on the zero or non-zero value of the
spin magnetization.  The transformations generated
by the action of $SU(3)$ generators were also
integrated into the Majorana geometrical picture.
Unlike qubits, the decomposition of single-qutrit
quantum gates in terms of radio-frequency pulses
is not straightforward and the Majorana sphere
representation provides a way to geometrically
describe these gates.  Close observations of the
dynamics of points representing a qutrit on the
Majorana sphere under the action of various
quantum gates were used to obtain the rf pulse
decompositions and basic single-qutrit gates were
experimentally implemented using NMR.

This work provides new insights into the intrinsic
features of a qutrit with the help of Majorana
representation, and introduces various NMR quantum
computing protocols for a single qutrit.  We think
these insights will be useful to all those engaged
in qutrit physics and particularly to those
researchers who are exploring the use of qutrits
as motifs for quantum computation.
\section*{Acknowledgements}
All experiments were performed on a 600
MHz Bruker Avance III FT-NMR spectrometer at the NMR 
Research Facility at IISER Mohali.
Arvind acknowledges funding from DST India
under Grant No.~EMR/2014/000297. KD acknowledges
funding
from DST India under Grant No.~EMR/2015/000556.
SD acknowledges University Grants Commission 
(UGC) India  for financial support.
\bibliographystyle{iopart-num} 
\providecommand{\newblock}{}

\end{document}